\documentclass[preprint,aps,  nofootinbib]{revtex4}
\usepackage{amssymb}
\usepackage{amsmath}
\usepackage{graphicx}

\newcommand\beq{ \begin{eqnarray} }
\newcommand\eeq{ \end{eqnarray} }

\begin{document}

\title{Shear Viscosity of a Non-Relativistic Conformal Gas in Two Dimensions}
\author{Jiunn-Wei Chen and Wen-Yu Wen}
\affiliation{Department of Physics and Center for Theoretical Sciences, National Taiwan
University, Taipei 10617}

\begin{abstract}
The shear viscosity, $\eta $, of a fermi gas with non-relativistic conformal
symmetry in two spatial dimensions is investigated. We find that $\eta /s$, $%
s$ being the entropy density, diverges as a gas of free particles in this
system. It is in contrast to the $\eta /s=1/4\pi $ found using
non-relativistic AdS/CFT correspondence, which requires a strongly
interacting CFT. It implies the unitary fermi gas in two spatial dimensions
is not likely to have a weakly interacting gravity dual.
\end{abstract}

\maketitle


\section{Motivation}

Recently the AdS/CFT correspondence, originally proposed for supersymmetric
conformal field theories \cite{Maldacena:1997re,Gubser:1998bc,Witten:1998qj}%
, has been conjectured to exist in non-relativistic conformal field theories
(NRCFT) \cite%
{Son:2008ye,Balasubramanian:2008dm,Goldberger:2008vg,Barbon:2008bg,Wen:2008hi}%
. One of the goals is to apply the tool to cold atomic systems in the
unitarity limit, where the two-body S-wave scattering length diverges (or
the two-body binding energy $B$ vanishes) and the system of two-component
fermions exhibits a non-relativistic conformal symmetry \cite{Mehen:1999nd} 
\footnote{%
If the system is bosonic or is fermionic but with more than two components
(e.g., with 2 spin and 2 isospin states), then the three-body interaction
can generate a scale to break the conformal symmetry \cite%
{Efimov:1970zz,Bedaque:1998kg}. For a system with two component (spin up and
down) fermions, the three-body interaction is derivatively coupled and is of
higher order by the Pauli exclusion principle.}. Later, the NR AdS/CFT
correspondence was generalized to finite temperature \cite%
{Herzog:2008wg,Maldacena:2008wh,Adams:2008wt}. In particular, a special kind
of black hole solution of type IIB was constructed using the Null Melvin
Twist technique. The theory was identified as the gravity dual of a $d=2$
NRCFT, $d$ being the number of spatial dimensions, at finite density and
finite temperature. The resulting shear viscosity $(\eta )$ to entropy
density $(s)$ ratio, $\eta /s$, is identical to $1/4\pi $ as in the
relativistic cases using AdS/CFT correspondence \cite%
{Policastro:2001yc,Policastro:2002se,Kovtun:2003wp}.

However, it is known that in $d=1$ and $2$, an attractive contact
interaction between two fermions will always give rise to a bound state.
Thus, zero binding energy implies a free system \cite%
{Landau,Nieto:2001bs,Nussinov,Nishida:2006eu}. In this paper, we demonstrate
this known result in the effective field theory (EFT) language. We conclude
that $\eta /s\rightarrow \infty $ in $d=2$ when $B=0$. This implies the
unitary fermi gas in $d=2$ is not likely to have a weakly interacting
gravity dual. It will be interesting to find some strongly interacting NRCFT
candidates in $d=2$ that might exhibit the NR AdS/CFT correspondence.

\section{The field theory approach}

For convenience, we use the EFT approach to compute the two-body scattering
amplitudes in various dimensions. This approach is equivalent to solving the
Schroedinger equation with a delta function potential. One can use a square
well potential to solve the Schroedinger equation then send the width of the
potential to zero such that the width does not break the conformal
invariance.

The leading order EFT Lagrangian in energy expansion for two-component,
non-relativistic fermions is \cite{Weinberg:1990rz,Kaplan:1998tg,Chen:1999tn}%
\begin{equation}
\mathcal{L}=\psi ^{\dagger }\left( i\partial _{t}+\frac{\nabla ^{2}}{2M}%
\right) \psi -C_{0}\left( \psi ^{\dagger }\psi \right) ^{2}\ ,
\end{equation}%
where four fermion contact interactions with derivatives are higher order
and are neglected. There is no particle pair creation in a non-relativistic
theory, so there is no contribution from the \textquotedblleft eye
diagrams.\textquotedblright\ The leading order two-body interaction through
the bubble diagrams shown in Fig. 1 gives rise to the scattering amplitude%

\begin{figure}[tbp]
\begin{center}
\includegraphics[height=2cm]{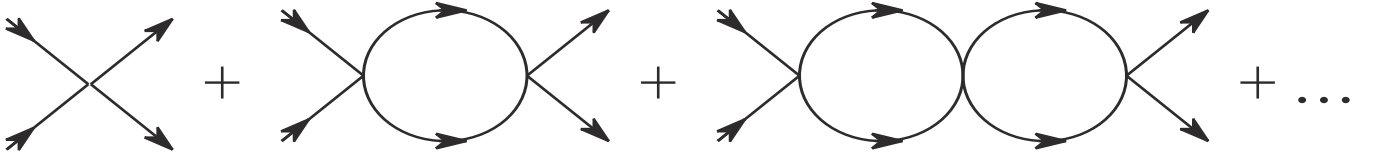}
\end{center}
\caption{Leading order diagrams for two fermion scattering.}
\end{figure}

\begin{equation}
i\mathcal{A}=-i\frac{C_{0}}{1-C_{0}I}=-i\frac{1}{1/C_{0}-I}\ ,  \label{A}
\end{equation}%
where $I$ denotes the loop integral. In the center-of-mass (CM) frame, the
system has energy $E$ and in dimensional regularization 
\begin{eqnarray}
I &=&-i\left( \frac{\mu }{2}\right) ^{d-D}\int \frac{d^{D+1}q}{\left( 2\pi
\right) ^{D+1}}\left( \frac{i}{\dfrac{E}{2}+q_{0}-\dfrac{\mathbf{q}^{2}}{2M}%
+i\epsilon }\right) \left( \frac{i}{\dfrac{E}{2}-q_{0}-\dfrac{\mathbf{q}^{2}%
}{2M}+i\epsilon }\right) \ ,  \notag \\
&=&\left( \frac{\mu }{2}\right) ^{d-D}\int \frac{d^{D}q}{\left( 2\pi \right)
^{D}}\frac{1}{E-\dfrac{\mathbf{q}^{2}}{M}+i\epsilon }\ ,  \notag \\
&=&-M\left( -ME-i\epsilon \right) ^{\frac{D-2}{2}}\Gamma \left( \frac{D-2}{2}%
\right) \left( \frac{\mu }{2}\right) ^{d-D}\left( 4\pi \right) ^{\frac{-D}{2}%
}\ ,  \label{B}
\end{eqnarray}%
where $d$ is the number of spatial dimensions and $D$ will be expanded
around $d$. If the interaction generates a bound state with bounding energy $%
B$, then $\mathcal{A}$ will have a pole at $E=-B$. Using Eqs. (\ref{A}) and (%
\ref{B}), this implies%
\begin{equation}
\frac{1}{C_{0}}=-M\left( MB\right) ^{\frac{D-2}{2}}\Gamma \left( \frac{D-2}{2%
}\right) \left( \frac{\mu }{2}\right) ^{d-D}\left( 4\pi \right) ^{\frac{-D}{2%
}}\ .
\end{equation}%
We will be interested in cases with $d=1,2,3$. 
\begin{equation}
\frac{1}{C_{0}}=\left\{ 
\begin{array}{l}
\dfrac{M\sqrt{MB}}{4\pi }\ ,\ d=3 \\ 
\dfrac{M}{2\pi \left( D-2\right) }+\frac{M}{4\pi }\left( \ln \left[ \dfrac{MB%
}{\mu ^{2}}\right] +\gamma _{E}\right) \ ,\ d=2 \\ 
-\frac{1}{2}\sqrt{\dfrac{M}{B}}\ ,\ d=1%
\end{array}%
\right.
\end{equation}%
When the system is tuned to have a bound state with zero binding energy ($%
B=0 $), we see that in $d=3$, $C_{0}\rightarrow \infty $ which is
corresponding to the unitarity limit where the two-body scattering length is
infinite. However, in $d=1$, $C_{0}\rightarrow 0$, which is the free case.
In $d=2$, $1/C_{0}$ has to absorb the $1/\left( D-2\right) $ pole and it
does not directly reflect the strength of the coupling. However, we can
analyze the scattering amplitude%
\begin{equation}
\mathcal{A}=\left\{ 
\begin{array}{l}
-\dfrac{4\pi }{M\sqrt{M}}\dfrac{1}{\sqrt{B}-i\sqrt{E}}\ ,\ d=3 \\ 
-\dfrac{4\pi }{M}\frac{1}{\left( \ln \left[ \dfrac{B}{E}\right] +i\dfrac{\pi 
}{2}\right) }\ ,\ d=2 \\ 
2\sqrt{\dfrac{B}{M}}\dfrac{i\sqrt{E}}{\sqrt{B}+i\sqrt{E}}\ ,\ d=1%
\end{array}%
\right.
\end{equation}%
By design, the amplitude $\mathcal{A}$ has a pole at $E=-B$ (the correct
limit for $B\rightarrow 0$ is to take $E=-B$ first then take $B\rightarrow 0$%
). We see that for $B=0$ and $E>0$, particles do not interact ($\mathcal{A}%
=0 $) in both $d=1$ or $d=2$. The same conclusion was obtained in \cite%
{Landau,Nieto:2001bs} by solving the Schroedinger equation.

The above analysis implies that shear viscosity $\eta \rightarrow \infty $
when $d=2$ (while $\eta $ is not defined in $d=1$). Note that the pole in
two-particle scattering amplitude at $E=0$ has no effect on $\eta $. This is
because $\eta $ reflects the time needed for a system to relax to thermal
equilibrium once it is perturbed away from equilibrium. However, $E=0$ in
the CM frame means there is no relative momentum between particles
scattering in any inertia frame. So there is no momentum rearrangement and
no relaxation to thermal equilibrium during the scattering. Thus, as far as
computing $\eta $ is concerned, the system is a free system and $\eta
\rightarrow \infty $. Since entropy density $s$ is finite for a free system, 
$\eta /s\rightarrow \infty $ for $d=2$ when $B=0$.

\section{Gravitational aspect}

In the gravity side, one might wonder if $\eta /s$ could have different
values for the free fermion limit and the $B=0$ limit. Just as in $d=3$,
both limits satisfy the same NR conformal symmetry but different boundary
conditions \cite{Son:2008ye}. In the following, however, we argue that in $%
d=2$, the two limits degenerate to the free fermion limit.

Let us recall the operator-field correspondence in NR AdS/CFT. We consider a
minimally coupled massive scalar field $\phi $ with mass $m$ propagating in
the following background of $d$ spatial dimensions, which exhibits a full
Schroedinger symmetry \cite{Son:2008ye} (see \cite{Duval:1990hj} for an
earlier work): 
\begin{equation}
ds^{2}=-\frac{2(dx^{+})^{2}}{z^{4}}+\frac{-2dx^{+}dx^{-}+dx^{i}dx^{i}+dz^{2}%
}{z^{2}}.
\end{equation}%
Here the two null-like Killing directions $\partial /\partial x^{+}$ and $%
\partial /\partial x^{-}$ are associated with energy $\omega $ and mass $M$
of the system and a discrete mass spectrum can be easily realized by making $%
x^{-}$\ periodic. Given a plane wave ansatz for a scalar field, 
\begin{equation}
\phi (x^{+},x^{-},x^{i},z)=e^{i\omega x^{+}+iMx^{-}+ik_{i}x^{i}}u(z),
\end{equation}%
one obtains two independent solutions \cite{Son:2008ye}: 
\begin{equation}
u_{\pm }=z^{d/2+1}K_{\pm \nu }(pz),\qquad p=\sqrt{\vec{k}^{2}-2M\omega }%
,\qquad \nu =\sqrt{m^{2}+2M^{2}+(\frac{d+2}{2})^{2}}.
\end{equation}%
For $0<\nu <1$, both solutions are renormalizable and the corresponding
operators have dimensions $\Delta _{\pm }=d/2+1\pm \nu $. In particular, one
is free to choose $\nu =d/2-1$ such that the operators have dimensions $d$
and $2$, respectively, corresponding to the dimension of the $\left( \psi
\psi \right) $ operator for free fermions and fermions at unitarity. Note
that, for $d=2$($\nu =0$), $u_{\pm }$ scales like $z^{2}$ and $z^{2}\ln
\left( z/z_{0}\right) $, where $z_{0}$ is some scale breaking the conformal
invariance \cite{Klebanov:1999tb}. Thus, only the $z^{2}$ solution is
allowed and we are left with a single picture of free fermions.

\section{Conclusions}

We have shown that for a system of two-component non-relativistic fermions
with $d=2$, as the two-body binding energy $B$ is tuned to be zero, $\eta
/s\rightarrow \infty $ as a free system. This implies the unitary fermi gas
in $d=2$ is not likely to have a weakly interacting gravity dual. It will be
interesting to find some strongly interacting NRCFT candidates in $d=2$ that
might exhibit the NR AdS/CFT correspondence.

\bigskip

\bigskip

We thank Allan Adams, Chris Herzog and Yuji Tachikawa for useful comments.
This work is supported by the NSC and NCTS of Taiwan.

\bibliographystyle{plain}
\bibliography{apssamp}

\end{document}